\documentclass[aps,twocolumn,groupedaddress,showpacs]{revtex4}
\usepackage{graphicx}
\usepackage{amsfonts}
\usepackage{amssymb}
\usepackage{amsmath}
\usepackage{wrapfig}

\begin{document}

\title
{Mean-field universality class induced by weak hyperbolic curvatures}

\author{Andrej~Gendiar$^1$, Michal Dani\v{s}ka$^1$,
Roman Kr\v{c}m\'{a}r$^{1,2}$, Tomotoshi Nishino$^2$}
\affiliation{
$^1$Institute of Physics, Slovak Academy of Sciences, SK-845~11, Bratislava,
Slovakia\\
$^2$Department of Physics, Graduate School of Science, Kobe University,
Kobe 657-8501, Japan}

\date{\today}

\begin{abstract}
Order-disorder phase transition of the ferromagnetic Ising model is investigated
on a series of two-dimensional lattices that have negative Gaussian curvatures. 
Exceptional lattice sites of coordination number seven are distributed on the
triangular lattice, where the typical distance between the nearest exceptional
sites is proportional to an integer parameter $n$. Thus, the corresponding
curvature is asymptotically proportional to $- n^{-2}_{~}$. Spontaneous magnetization
and specific heat are calculated by means of the corner transfer matrix
renormalization group method. For all the finite $n$ cases, we observe the
mean-field-like phase transition. It is confirmed that the entanglement entropy
at the transition temperature is linear in $( c / 6 ) \ln n$, where $c = 1 / 2$ is the
central charge of the Ising model. The fact agrees with the presence of the
typical length scale $n$ being proportional to the curvature radius.
\end{abstract}


\maketitle

\section{Introduction}

Quantum and statistical phenomena under non-Euclidean geometry have 
been attracting research interests in a number of physical systems.
For example, experiments on magnetic
nano-structures~\cite{experiment1,experiment2,experiment3} 
have been performed in connection with soft materials exhibiting
a conical geometry~\cite{cn}. One can list investigations on the 
quantum gravity~\cite{q-gravity1,q-gravity2}, lattice dislocations
of the solid-state crystals, complex networks~\cite{cnet1,cnet2}
such as neural systems and complicated web connections. 

The focus of the present analysis lies in classical lattice spin models,
for which the thermal property is influenced by the non-flatness of the
underlying lattice. As typical examples, phase transitions on regular
two-dimensional hyperbolic lattices have been studied for
the Ising model~\cite{Shima,hctmrg-Ising-5-4,Sakaniwa},
the $q$-state clock models~\cite{hctmrg-clock-5-4,Baek-clock}, 
the $XY$-model~\cite{XY-model}, and the frustrated
$J_1^{~}$-$J_2^{~}$ Ising model~\cite{hctmrg-J1J2}. 
In these studies, the hyperbolic lattices are constructed by means of the
tessellation of regular polygons with $p$ sides, where the
coordination number, $q$, satisfies the hyperbolic condition 
$( p - 2 ) ( q - 2 ) > 4$. Such uniform hyperbolic lattices are
conventionally called the $( p, q )$ lattice, and their Hausdorff
dimension is infinite~\cite{Mosseri,hctmrg-Ising-3-q}. 
The observed phase transitions in numerical studies qualitatively 
agree with the mean-field approximation. In particular, the second-order
phase transition with the Landau mean-field universality has been 
observed for the Ising model~\cite{hctmrg-Ising-p-4}. 
The origin of the mean-field behavior has not been clarified yet.

A key feature of the phase transitions on the hyperbolic $( p, q )$ lattices
is that the correlation length remains finite even at the transition 
temperature~\cite{hctmrg-tr-mat,hctmrg-Ising-3-q}.
The fact suggests the presence of an inherent length in each $( p, q )$ 
lattice, the length which violates the realization of the scale invariance
at criticality. One can conjecture that the length is related to the curvature 
radius $r = 1 / \sqrt{- K}$, where $K$ is the Gaussian curvature 
on a hyperbolic plane. As long as we observe $( p, q )$ lattices only, 
it is non-trivial to confirm this conjecture, since the radius $r$
is of the order of the lattice constant even on the $( 3, 7 )$ lattice,
which is less curved than the cases with $p = 3$ and $q \ge 8$.

In this article we propose a way to construct a series of two-dimensional lattices 
formed by tessellation of the triangles ($p = 3$), where the absolute value
of the {\it averaged} curvature $K$ is smaller than the $( 3, 7 )$ lattice. 
Such a slightly curved lattice is obtained by distributing exceptional lattice
sites of the coordination number seven within the triangular lattice in such
manner that the typical distance between nearest exceptional sites is
proportional to an integer parameter $n$. Under such construction, the
corresponding curvature $K$ is asymptotically proportional to 
$n^{-2}_{~}$, and the radius $r$ is proportional to $n$. 
We calculate thermodynamic properties of the Ising model on the series 
of weakly curved lattices by means of the corner transfer matrix renormalization 
group (CTMRG) method~\cite{ctmrg-tn}, which is based on Baxter's corner 
transfer matrix (CTM) scheme~\cite{Baxter}. As a quantity that captures
the length scale, we focus on the entanglement entropy at the
transition temperature. As we show in the following, the entropy scales 
as $( c / 6 ) \ln n$, where $c = 1 / 2$ is the central charge of the Ising model. 
The fact supports the conjecture on the presence of a finite length scale,
which is related to the curvature radius at the transition temperature.

The paper is organized as follows. Section~II is devoted to the construction of
the series of slightly curved lattices constructed by the triangular tessellation. 
We evaluate the curvature $K$ and the corresponding radius $r$ in two ways, 
one is the averaged curvature on the whole lattice, and the other one is obtained
around the center of the lattice. Both evaluations are in agreement with the
asymptotic form of the curvature, which is proportional to $n^{-2}_{~}$. In Sec.~III
we derive thermodynamic quantities by means of the CTMRG method, which is modified
for the series of the lattices. We obtain the spontaneous magnetization, specific
heat, and the entanglement entropy. The obtained results are summarized in the
last section.

\section{Hyperbolic Lattices}

Throughout this article we consider the Ising model on two-dimensional
lattices being either flat or negatively curved. The Hamiltonian is written as
\begin{equation}
H\{ \sigma \} = 
- J \sum_{\langle i, j \rangle}^{~} \sigma_i^{~} \sigma_j^{~}
- h \sum_{\langle i \rangle}^{~} \sigma_i^{~}\, ,
\label{Hm}
\end{equation}
where $\sigma_i^{~} = \pm 1$ represents the Ising spin on the lattice site
labeled by $i$, and the notation $\langle i, j \rangle$ denotes the nearest
neighboring sites. We assume that the interaction is ferromagnetic ($J > 0$)
and a constant magnetic field $h$ is acting equally on each spin site. We
always keep in mind the possibility of obtaining the partition function 
\begin{equation}
Z = \sum_{ \{\sigma\} }^{~} 
          \exp\left[ - \frac{ 1 }{ k_{\rm B}^{~} T } \, H\{ \sigma \} \right] 
\end{equation}
numerically, by means of the CTMRG method~\cite{hctmrg-Ising-3-q,ctmrg-tn}.
Under these requirements, we have chosen the following candidates of the
curved lattices.

\begin{figure}[tb]
\centerline{
\includegraphics[width=0.22\textwidth,clip]{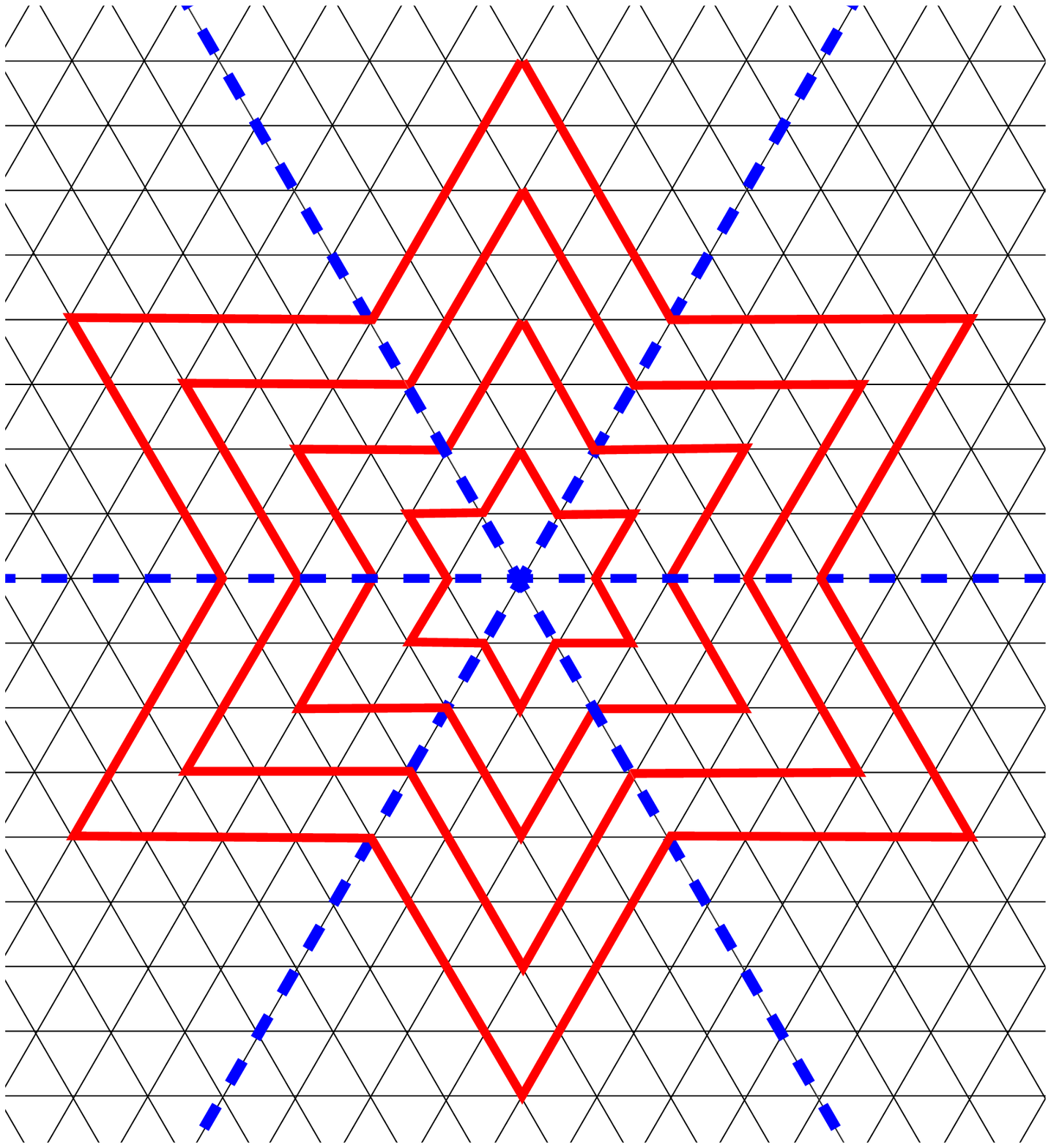}
\includegraphics[width=0.250\textwidth,clip]{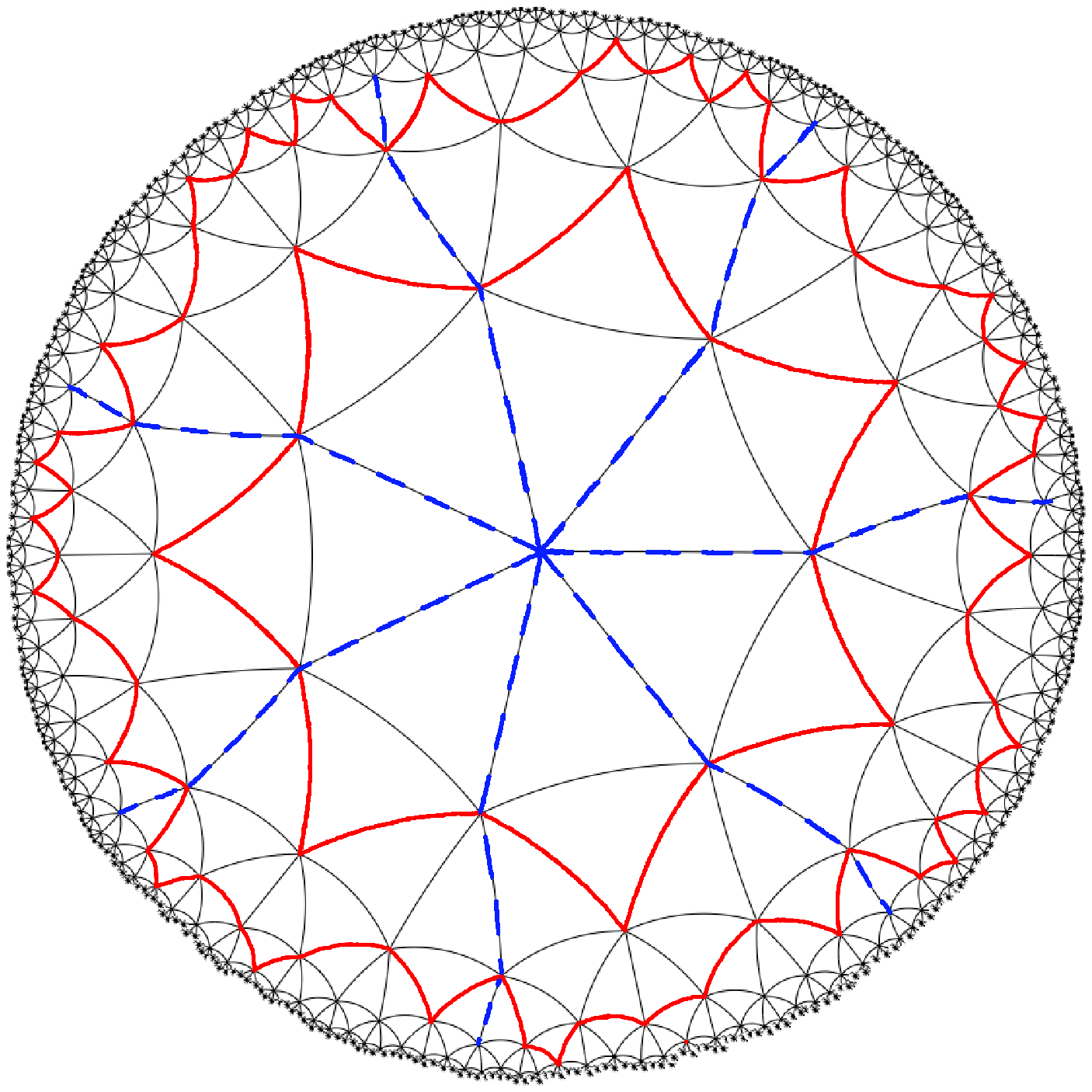}}
\caption{(Color online) The flat $( 3, 6 )$ lattice on the left and hyperbolic 
$( 3, 7 )$ one on the right. The blue dashed lines divide each lattice into identical 
areas, the corners, meeting at the center. The star-shaped area depicted 
by the thick curves in red show the finite areas of the lattices after $M$ steps of 
iterative extensions in Eq.~(6).
}
\label{fig1}
\end{figure}

Let us start from the $( p, q )$ lattice, which is a tessellation of regular 
polygons with $p$ sides, where the coordination number around each site
is $q$. We restrict ourselves to the case of $p = 3$ throughout this article.
Figure~\ref{fig1} shows two examples, a triangular (planar) $( 3, 6 )$ lattice 
and a hyperbolic $( 3, 7 )$ lattice. The latter is drawn inside the so-called
Poincare disk because the $( 3, 7 )$ lattice has a non-Euclidean geometry. 
Although the triangles inside the Poincare disk are deformed and shrunk 
toward the border of the circle, which corresponds to the infinity, the
interaction coupling $J$ remains constant everywhere, and so does the actual
sizes of the triangles. The blue dashed lines divide the lattice into $q$
equivalent parts, ${\cal C}$, which are called the corners. 

The $( 3, q )$ lattice can be constructed by means of recursive
extensions~\cite{hctmrg-Ising-3-q}.
In order to simplify the discussion, we start from the $( 3, 6 )$ lattice.  
The smallest unit we consider is not a equilateral triangle. Instead, we chose
a rhombus ${\cal W}$ consisting of two adjacent equilateral triangles. We
introduce other two objects, parallelograms (or stripes) ${\cal L}_M^{~}$
and ${\cal R}_M^{~}$ created by joining $M$ number of rhombi ${\cal W}$
in one direction. Let us write such joining process by using formal recursive
equations
\begin{eqnarray}
{\cal L}_{M+1}^{~} &=& {\cal W} \, {\cal L}_M^{~} \, , \nonumber\\
{\cal R}_{M+1}^{~} &=& {\cal W} \, {\cal R}_M^{~} \, ,
\label{hrtm36}
\end{eqnarray}
initiated from ${\cal L}_1^{~} = {\cal R}_1^{~} = {\cal W}$. These 
products on the right-hand side represent the joining of parts in a pictorial 
(or diagrammatic) manner. We also need to introduce another extended rhombus
${\cal C}_M^{~}$ of the size $M$ by $M$ satisfying the formal joining relation
\begin{equation}
{\cal C}_{M+1}^{~} = {\cal W} \, {\cal L}_M^{~} {\cal C}_M^{~} {\cal R}_M^{~}
\label{ctm36}
\end{equation}
starting from ${\cal C}_1^{~} = {\cal W}$. We often call ${\cal C}_M^{~}$
a corner. Hence, we can consider a star-shaped area proportional to the size
$M$ that is constructed by joining six corners that are formally represented
as $\left( {\cal C}_M^{~} \right)^6_{~}$. The red lines in Fig.~\ref{fig1}
(left) bound the areas for the cases $1 \le M \le 4$. Repeating this extension
processes, the star-shaped area of an arbitrary size can be obtained on the
$( 3, 6 )$ lattice. The total number of the lattice sites in
$\left( {\cal C}_M^{~} \right)^6_{~}$ is $6 ( M + 1 ) M + 1$.

A slight modification of the extension processes in Eqs.~\eqref{hrtm36} and
\eqref{ctm36} enables us to construct the hyperbolic $( 3, 7 )$ lattice,
shown in the right side of Fig.~\ref{fig1}.  In this case, the extensions are
formally written as
\begin{eqnarray}
\label{tm37}
{\cal L}_{M+1}^{~} &=& {\cal W} \, {\cal L}_M^{~} \, {\cal C}_M^{~} \, , \nonumber\\
{\cal R}_{M+1}^{~} &=& {\cal W} \, {\cal C}_M^{~} \, {\cal R}_M^{~} \, , \\
{\cal C}_{M+1}^{~} &=& 
{\cal W} \, {\cal L}_M^{~} \, \left( {\cal C}_M^{~} \right)^2_{~} \, {\cal R}_M^{~}  \, , \nonumber
\end{eqnarray}
where the details can be found in Ref.~\cite{hctmrg-Ising-3-q}. Notice that
the extension to the hyperbolic $( 3, 7 )$ lattice also follows a recursive
construction. Compared with the extension process in Eqs.~\eqref{hrtm36} and
\eqref{ctm36} of the $( 3, 6 )$ lattice, the right-hand sides of 
Eq.~\eqref{tm37} contain an extra corner ${\cal C}_M^{~}$, and this
insertion realizes the coordination number seven within the whole lattice. 
The areas on the right side of Fig.~\ref{fig1}, bordered by the red lines,
correspond to the `star-shaped' lattices $\left( {\cal C}_1^{~} \right)^7_{~}$
and $\left( {\cal C}_2^{~} \right)^7_{~}$. On the $( 3, 7 )$ lattice, the
total number of the lattice sites grows exponentially with $M$~\cite{M37}.

Among the $( 3, q )$ lattices satisfying the hyperbolic condition 
$q > 6$, the $( 3, 7 )$ lattice exhibits the least curvature in the sense
that the absolute value of its curvature, $| K |$, is the smallest. The
curvature radius $r = 1 / \sqrt{ - K }\approx 0.917$ is already of the
order of the lattice constant~\cite{hctmrg-Ising-3-q}, in contrast to
$r = \infty$ in the $( 3, 6 )$ lattice. In this respect, the $( 3, 7 )$ 
lattice is `too far' from the $( 3, 6 )$ lattice. We have to construct such
lattices that have the curvature radii in between, i.e., $0.917 < r < \infty$,
in order to quantify the effect of the non-zero curvature to the 
order-disorder phase transition. We, therefore, consider such a lattice
that consists of triangles, and the lattice sites contain a mixture of
the coordination numbers six and seven. As the number of the lattice sites
with the coordination number seven decreases, such a mixed lattice approaches
the flat triangular $( 3, 6 )$ lattice. We use the term `exceptional' lattice
site for such sites that have the coordination number seven.

There are many sequential methods to generate mixed lattices. 
We have chosen the following extension scheme
\begin{eqnarray}
\label{tm3qn}
{\cal L}_{M+1}^{~} &=& {\cal W} \, {\cal L}_M^{~}  \, , \nonumber\\
{\cal R}_{M+1}^{~} &=& {\cal W} \, {\cal R}_M^{~} \, , \\
{\cal C}_{M+1}^{~} &=& \left\{
\begin{array}{ll}
{\cal W} \, {\cal L}_M^{~} \left( {\cal C}_M^{~} \right)^2_{~} {\cal R}_M^{~}
\qquad ( {\rm at} ~ {\rm every} ~ n^{\rm th} ~{\rm step} ), \nonumber\\ 
{\cal W} \, {\cal L}_M^{~} \,\,\, {\cal C}_M^{~} \,\,\,\, {\cal R}_M^{~}
\qquad \,\, ( {\rm otherwise} )
\end{array}
\right.
\end{eqnarray}
to analyze the property of the Ising model on this lattice.
These processes are almost the same as the extension scheme
in Eqs.~\eqref{hrtm36} and \eqref{ctm36} for the $( 3, 6 )$
lattice, but when $M$ is a multiple of an integer parameter 
$n$, we insert an additional corner ${\cal C}_M^{~}$ in the
extension process from ${\cal C}_M^{~}$ to ${\cal C}_{M+1}^{~}$.
This process adds the exceptional lattice site with the
coordination number seven whenever $(M \mod n)=0$.
Note that we used the extension
process of ${\cal L}_M^{~}$ and ${\cal R}_M^{~}$ as in
Eq.~\eqref{hrtm36}. This restriction keeps the corner
${\cal C}_M^{~}$ symmetric to the spatial inversion, the
property which is convenient for numerical calculations
by the CTMRG method. On the other hand, this simplification
introduces a slight inhomogeneity to the lattice, which
should be considered carefully. 

\begin{figure}[tb]
\centerline{
\includegraphics[width=0.24\textwidth,clip]{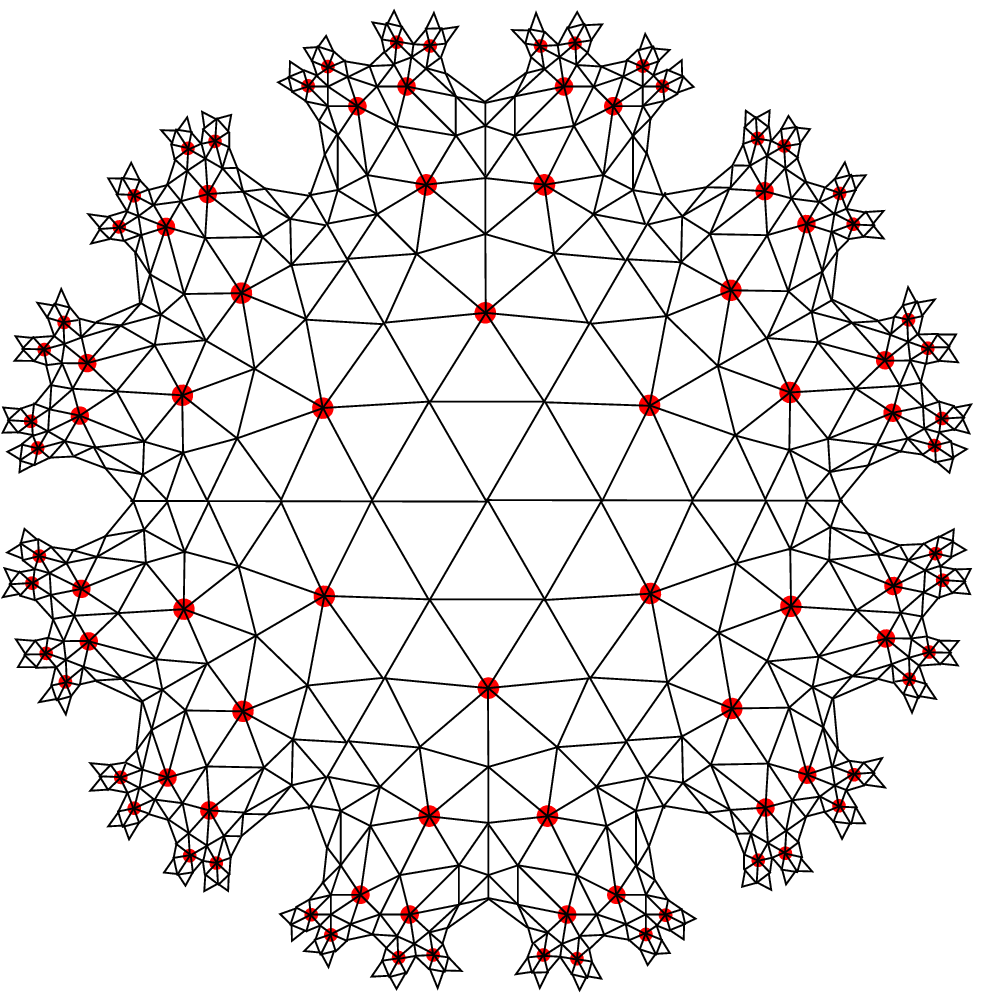}
\includegraphics[width=0.24\textwidth,clip]{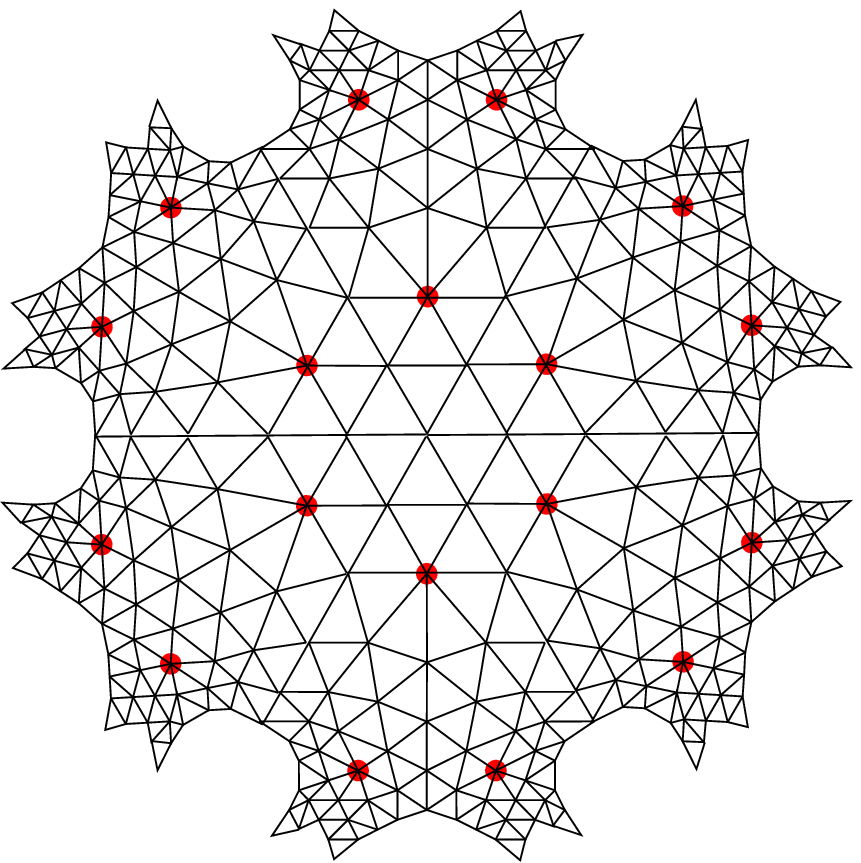}}
\caption{(Color online) The two typical lattice geometries for
$n=1$ (left) and $n=2$ (right) for the size $M = 5$. The filled
circles denote the exceptional lattice sites with the coordination
number seven, and number of the exceptional sites is $90$ in the left
and $18$ in the right, c.f. Eq.\eqref{SnM}.}
\label{fig2}
\end{figure}

Whenever we obtain the extended corer ${\cal C}_{M+1}^{~}$
shown as Eq.~\eqref{tm3qn}, we consider the joined lattice area
made of the six corners which can be formally represented
as $( {\cal C}_{M+1}^{~} )^6_{~}$. 
Figure~\ref{fig2} shows the two examples of such `star-shaped'
regions for $M = 5$ in the cases when $n=1$ (left) and $n=2$
(right). The filled dots (in red) emphasize those exceptional
lattice sites, where the additional corners have been inserted. 

\subsection{Coordination number}

Looking at the extension process in Eq.~\eqref{tm3qn}, one finds that
total number of the lattice sites ${\cal N}_n^{~}( M )$ exponentially
increases with $M$ for arbitrary finite $n$. When $n$ is a multiple of
$M$, this counting is easily performed by a recursive formula shown in
Appendix, and generalization to the arbitrary $n$ is straightforward. 
Having counted the total number of the lattice sites in the whole lattice
area $({\cal C}_{M})^6$ created by Eq.~\eqref{tm3qn}, we obtain
\begin{equation}
{\cal N}_n^{~}( M ) = 1+12\sum\limits_{j=1}^{M}
j\, 2^{k_n^{~}(nk_n^{~}(M,1)+n, \, j)}\, ,
\end{equation}
where we introduced a double-nested greatest integer (floor) function
in the exponent; the floor function has the following form
\begin{equation}
k_n(m,j)=\left\lfloor\frac{m-j}{n}\right\rfloor\equiv\max\left\{i\in{\mathbb Z}
\ \vert\ i\leq\frac{m-j}{n}\right\}\, .
\end{equation}
In the same manner, we can obtain the number of the exceptional
sites
\begin{equation}
\label{SnM}
{\cal S}_n(M)=6\left[ 2^{k_n(M,1)} - 1 \right]
\end{equation}
for any set of $n$ and $M$. This number is consistent with the cases
shown in Fig.~\ref{fig2}, where ${\cal S}_{1}(5) = 90$ on the left and
${\cal S}_{2}(5) = 18$ on the right.

Considering the asymptotic limit $M \rightarrow \infty$, the ratio between
${\cal S}_n(M)$ and ${\cal N}_n(M)$ leads to the average density
of the exceptional sites
\begin{equation}
\lim\limits_{M \rightarrow \infty}^{~} \frac{{\cal S}_n(M)}{{\cal N}_n(M)}
= \frac{1}{2n(3n+1)} \, .
\end{equation}
For sufficiently large $n$, the density becomes proportional to $n^{-2}_{~}$.
For the brevity, we introduce the averaged coordination number
\begin{equation}
q_n^{~} = 6 + \frac{1}{2n(3n+1)} \, .
\label{qn}
\end{equation}
Note that $q_{\infty}^{~} = 6$ is the coordination number of the $( 3, 6 )$
lattice~\cite{g36}. Using the notation $q_n^{~}$ thus defined, we denote the
lattice constructed by Eq.~\eqref{tm3qn} as the $( 3, q_n^{~} )$ lattice.

Length of the system lattice border, ${\cal P}_n^{~}(M)$, is another essential
quantity that characterizes the geometry of the $( 3, q_n^{~} )$ lattice. 
The analytic formula of ${\cal P}_n^{~}(M)$ can be obtained as
\begin{equation}
{\cal P}_{n}^{~}(M)=12\left[ M - n k_n^{~}(M,1) + n \sum\limits_{j=1}^{k_n^{~}(M,1)} 2^j_{~} \right] \, ,
\end{equation}
where a simple derivation is presented in the Appendix.
It should be noted that the ratio of the boundary sites to the total number
of the lattice sites in the asymptotic limit
\begin{equation}
\lim\limits_{M \rightarrow \infty} \frac{{\cal P}_n(M)}{{\cal N}_n(M)} = 
\frac{2}{3 n + 1}
\end{equation}
is finite and inversely proportional to $n^{-1}$. Such a dominance of the boundary
sites over all lattice sites is a characteristic feature of hyperbolic lattices.
Our research target, the thermodynamic property of the system at the center of
the $( 3, q_n^{~} )$ lattice, is thus surrounded by a wide system boundary which
increases exponentially.

\subsection{Averaged curvature}

Now, let us focus our attention to the curvature of the $( 3, q_n^{~} )$ lattice.
If one looks at a small region that does not contain any exceptional lattice sites, 
the region is identical to the $( 3, 6 )$ lattice as long as the connection of the
lattice sites is concerned. The hyperbolic nature of the $( 3, q_n^{~} )$ lattice
arises from the presence of the exceptional lattice sites which are distributed in
a sparse manner. Thus, when we consider the curvature of the $( 3, q_n^{~} )$
lattice, we have to take a certain average over the system. Apparently such an
averaged curvature is dependent to the parameter $n$, and we write it as $K_n^{~}$
in the following. Roughly speaking, $K_n^{~}$ should be proportional to
$n^{-2}_{~}$ since the natural scale of the $( 3, q_n^{~} )$ lattice is given
by $n$. We evaluate the averaged curvature by
\begin{equation}
K_n^{~} = - r_n^{-2} \, ,
\label{Kn}
\end{equation}
where $r_n^{~}$ is the corresponding curvature radius using a geometrical
formula~\cite{Mosseri}
\begin{equation}
\cosh \frac{1}{2 r_n^{~}} =
\frac {\displaystyle \cos \frac{\pi}{3} }{\displaystyle \sin \frac{\pi}{q_n^{~}} }
\label{ModMos}
\end{equation}
on a hyperbolic triangle that consists the $( 3, q_n^{~} )$ lattice.
We have chosen the lattice constant as the unit of the length. Substituting
the asymptotic expression $q_n=6+1/6n^2$ from Eq.~\eqref{qn} into
Eq.~\eqref{ModMos}, we obtain 
\begin{equation}
K_n^{~} \, \sim \, - \frac{2}{3\pi} n^{-2}
\label{Kqn}
\end{equation}
with the dominant coefficient $2/3\pi \approx 0.212$ for large $n$.

Complementing the evaluation of the averaged curvature, we relate the
length of the lattice boundary, ${\cal P}_n(M)$, to the curvature radius
$r_n$ on a hyperbolic plane
\begin{equation}
{\cal P}_n(M) \propto 2 \pi \, \sinh \frac{M}{r_n} \, .
\label{LnM}
\end{equation}
Using Eq.~\eqref{Kn} and taking the limit $M\to\infty$, we obtain
\begin{equation}
K_n^{~} \, \sim \, - ( \ln \, 2 )^2_{~} \, n^{-2}_{~}  \, ,
\end{equation}
where the prefactor $( \ln \, 2 )^2_{~}\approx 0.48$. To summarize, we have
evaluated the averaged curvature on the $( 3, q_n^{~} )$ plane in two ways,
and both of them lead to $K_n \propto - n^{-2}_{~}$.

\section{Numerical results}

In this section we study the phase transition of the Ising model on the sequence
of the non-Euclidean $( 3, q_n^{~} )$ lattices, in particular,
\begin{equation}
( 3, q_1^{~} ), \quad ( 3, q_2^{~} ), \quad ( 3, q_3^{~} ), \quad
\cdots, \quad ( 3, q_{\infty}^{~} )\, .
\end{equation}
The Hamiltonian is given by Eq.~\eqref{Hm},
and without loss of generality, the coupling constant $J$ and the Boltzmann
constant $k_{\rm B}^{~}$ are chosen to be unity. All thermodynamic functions
are considered in dimensionless units. Since the elementary unit of
$( 3, q_n^{~} )$ lattice is the rhombus-shaped ${\cal W}$, it is natural
to attribute the Boltzmann weight to each ${\cal W}$. Suppose that the
Ising spins $\sigma_i^{~}$, $\sigma_j^{~}$, $\sigma_k^{~}$, and $\sigma_l^{~}$
are placed on the corners of the rhombus. The corresponding Boltzmann weight
${\cal W}$ is given by
\begin{eqnarray}
\nonumber
&{\cal W} ( \sigma_i^{~} \sigma_j^{~} \sigma_k^{~} \sigma_l^{~} )
 = \exp \bigg\{\frac{\displaystyle\strut J}{\displaystyle\strut  2 k_{\rm B}^{~} T}
   \left( \sigma_i^{~} \sigma_j^{~} + \sigma_j^{~} \sigma_k^{~}
   + \sigma_k^{~} \sigma_l^{~} \right. \\
\nonumber
& + \left. \sigma_l^{~} \sigma_i^{~} + 2 \sigma_j^{~} \sigma_l^{~} \right)
   +  \frac{\displaystyle\strut  h}{\displaystyle\strut  6 k_{\rm B}^{~} T} \left( \sigma_i^{~}
   + 2 \sigma_j^{~} + \xi \sigma_k^{~} + 2 \sigma_l^{~} \right)
\bigg\} \, , \\
&
\end{eqnarray}
where the diagonal interaction acts between the spins $\sigma_j^{~} $ and
$\sigma_l^{~} $. The pre-factor $\xi$ is normally unity, and is set to zero
when over-counting of interaction with external field $h$ happens at each
exceptional lattice point. Most of the numerical calculations are performed
under $h = 0$ in the following; the only exception is when we observe the
magnetic response at the transition temperature.

Taking the tensor product among weights ${\cal W}$, one can gradually expand
the size of the Boltzmann weights ${\cal L}_M^{~}$  and ${\cal R}_M^{~}$.
These weights are called the half-row transfer matrices. Analogously, the
expanding weight ${\cal C}_M^{~}$ is called the corner transfer matrix. 
The procedure of obtaining the transfer matrices represent a generalized
version of the CTMRG method applied to the $( 3, 7 )$ lattice which is
studied in detail in Ref.~\cite{hctmrg-Ising-3-q}. Consequently, the
`reduced' density matrix is a partial trace of the corner transfer
matrices
\begin{equation}
\rho_n^{~}(M) = {\rm Tr}^{\prime}\left[{C_M^{~}}\right]^6_{~} \, ,
\end{equation}
where we explicitly include the parameter $n$ in $\rho_n^{~}(M)$.
In the following we omit the size dependence on $M$ of the reduced
density matrix to simplify the formulae. Taking the complete trace
of the reduced density matrix leads to the partition function~\cite{ctmrg-tn}
\begin{equation}
Z_{n}^{~} = {\rm Tr} \, \rho_{n}^{~}
\label{pf}
\end{equation}
of the star-shaped lattice area.

In our numerical calculations by CTMRG, we keep up to $m=200$ block spin
states~\cite{ctmrg-tn,hctmrg-Ising-5-4,hctmrg-Ising-p-4}, where we have
confirmed that all the data are converged with respect to $m$. As the system
size $M$ increases, ${\cal C}_M^{~}$ approaches its thermodynamic limit during the
numerical calculations. Note that ${\cal C}_M^{~}$ possesses a minor 
dependence on $M$, since we keep inserting of the exceptional lattice sites
at every $n^{\rm th}$ extension step in accord with Eq.~\eqref{tm3qn}.
We can either consider the cases where $M$ is multiple of $n$ or take the
average among the minor fluctuations. There is, however, no qualitative
difference in the two choices, and we have chosen the latter one. 
It should be noted that we focus on the thermodynamic quantity deep inside
the system, and discard those phenomena near the system boundary, as we have
considered in the previous studies.~\cite{hctmrg-Ising-3-q}

\begin{figure}[tb]
\centerline{\includegraphics[width=0.5\textwidth,clip]{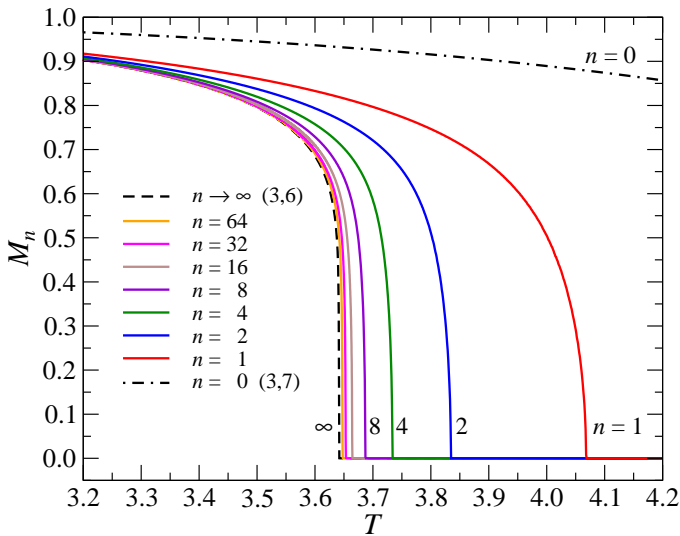}}
\caption{(Color online) Temperature dependence of the spontaneous
magnetization $M_n^{~}( T )$ on the $( 3, q_n^{~} )$ and $( 3, 7 )$
lattices~\cite{hctmrg-Ising-3-q}.}
\label{fig3}
\end{figure}

Spontaneous magnetization provides information in the ordered phase.
Figure~\ref{fig3} displays the temperature dependence of the bulk
spontaneous magnetization
\begin{equation}
M_n^{~}( T ) =  \frac{{\rm Tr} \left( \sigma_c^{~} \rho_n^{~} \right)}{{\rm Tr}
\, \rho_n^{~}} 
\label{mag}
\end{equation}
at $h = 0$, the value which measures the average polarization of the spin
$\sigma_c^{~}$ at the center of the lattice system. For comparison, we also
show the magnetization on the flat $( 3, 6 )$ lattice, denoted by $n \to \infty$,
as well as on the hyperbolic $( 3, 7 )$ lattice, denoted by $n = 0$~\cite{hctmrg-Ising-3-q};
we use the analogous notation (by the subscript $n$) for other thermodynamic
quantities. The phase transition temperature $T_n^{~}$ monotonously decreases
with $n$ and approaches the analytically known values $T_{\infty}^{~} =
4 / \ln 3 \sim 3.64096$~\cite{Baxter} on the flat $( 3, 6 )$ lattice. Roughly
speaking, the difference $T_n^{~} - T_{\infty}^{~}$ is inversely proportional
to $n$. 

\begin{figure}[tb]
\centerline{\includegraphics[width=0.5\textwidth,clip]{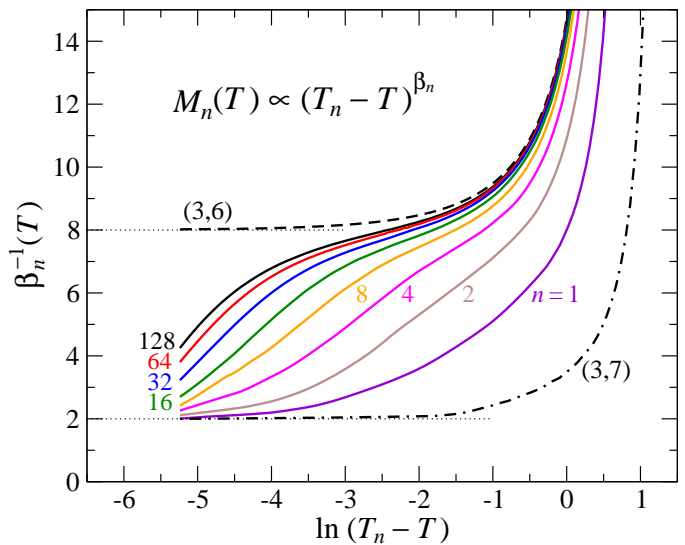}}
\caption{(Color online) Inverse of the effective magnetic exponent
$\beta_n^{~}( T )$ as a function of the logarithmic distance from the
transition temperature.}
\label{fig4}
\end{figure}

Just below the transition temperature the power-law behavior
\begin{equation}
M_n^{~}( T ) \propto {\left(T_n^{~} - T \right)}^{\beta_n}_{~} \, ,
\end{equation}
is expected. In order to detect the magnetic exponent $\beta_n^{~}$
from the numerically calculated $M_n^{~}( T )$, we use the derivative
\begin{equation}
\beta_n^{~}( T ) = \frac{\partial \ln M_n^{~}( T )}{\partial \ln \left(T_n^{~} - T \right)} \, ,
\label{beff}
\end{equation}
within the ferromagnetic ordered phase $T \leq T_n^{~}$. 
Figure~\ref{fig4} shows $\beta_n^{~}( T )$ thus obtained. 
When $T_n^{~} - T$ is relatively large, $\beta_n^{~}( T )$
follows the Ising universality where $\beta=\frac{1}{8}$,
however, in the neighborhood of the transition temperature $T_n^{~}$,
the magnetic exponent $\beta_n^{~}$ for finite $n$ increases
and tends to $\beta_n=\frac{1}{2}$, the value which represents
the mean-field universality class.

\begin{figure}[tb]
\centerline{\includegraphics[width=0.5\textwidth,clip]{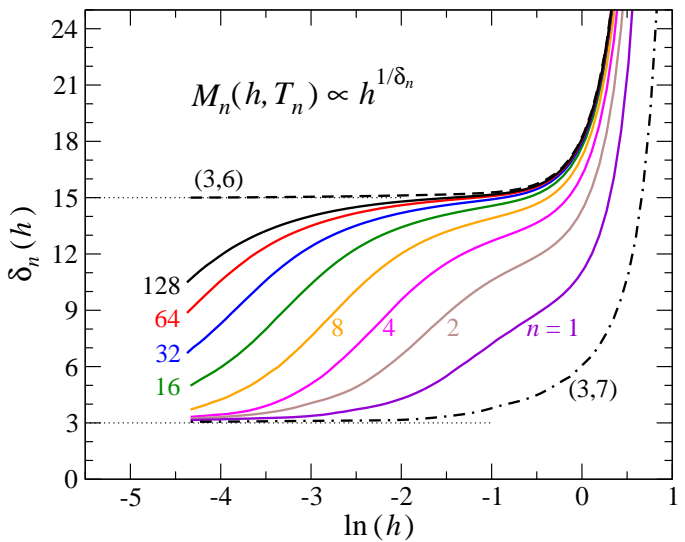}}
\caption{(Color online) Effective exponent $\delta_n^{~}( h )$ with
respect to the logarithm of the external magnetic field at the phase
transition temperature $T_n^{~}$.}
\label{fig5}
\end{figure}

In addition, we studied the exponent $\delta$ which is associated
with the response of the magnetization to a uniform magnetic field
$h$ at the phase transition temperature $T_n^{~}$, which obeys the
scaling
\begin{equation}
M_n^{~}( h, T_n^{~} ) \propto h^{1/\delta_n}_{~} 
\end{equation}
on the planar lattice. Figure~\ref{fig5} shows the effective critical
exponent
\begin{equation}
\delta_n^{~}( h ) =
      {\left[
         \frac{\partial \ln M_n^{~}( h, T_n^{~} )}{\partial \ln h }
       \right]}^{-1}
\label{deff}
\end{equation}
in the limit $h\to0$. The observed behavior qualitatively agrees with
that of the magnetic exponent $\beta$ depicted in Fig.~\ref{fig4};
the Ising universality $\delta=15$ is recovered for the ($3,6$) lattice
only. It is obvious that the effective exponent $\delta_n^{~}( h )$ 
deviates from the Ising one when the external field becomes small, and
it again approaches the mean-field value $\delta_n^{~}(h\to0)=3$ for
any finite $n$.

\begin{figure}[tb]
\centerline{\includegraphics[width=0.5\textwidth,clip]{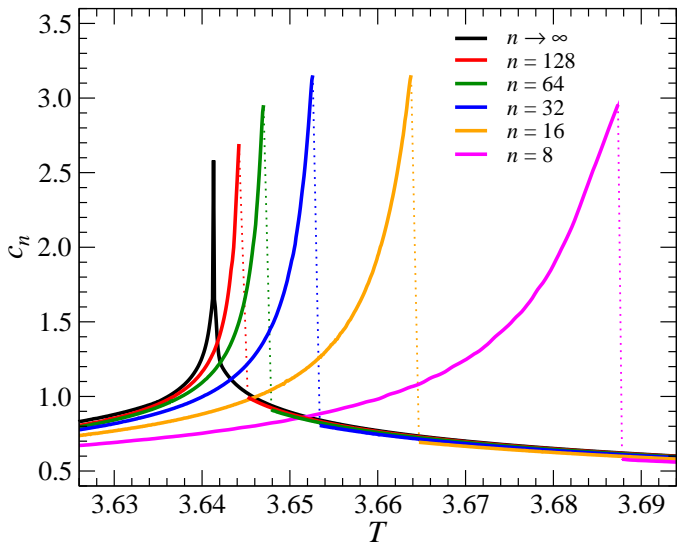}}
\caption{(Color online) The specific heat on the $( 3, q_n^{~} )$ lattice.}
\label{fig6}
\end{figure}

The internal energy at the center of the system is represented as
\begin{equation}
U_n^{~}( T ) =  -J\, \frac{{\rm Tr} \left( \sigma_c^{~} \sigma_{c'}^{~} \, \rho_{n}^{~} \right)}{
{\rm Tr} \, \rho_n^{~}} \, , 
\label{spc}
\end{equation}
where $\sigma_c^{~}$ and $\sigma_{c'}^{~}$ are, respectively, the spin at
the center of the system and a neighboring one. Figure~\ref{fig6} shows the
specific heat $c_n^{~}( T )$, which is obtained by taking the numerical
derivative of $U_n^{~}( T )$ with respect to the temperature $T$. The maxima
of the specific heat for large $n$ are not obtained precisely, because
$U_n^{~}( T )$ around $T = T_n^{~}$ is very sensitive to a tiny numerical error.
The discontinuity in $c_n^{~}( T )$ for finite $n$ supports the fact that the
transition is of the mean-field nature. Note that small differences of the
specific heat, $c_n^{~}( T )$, in the disordered region $T \ge T_n^{~}$ for
various $n$ is close to $c_{\infty}^{~}( T )$ on the flat $( 3, 6 )$
lattice. This suggests a transient behavior from the Ising universality to
the mean-field one which happens within the disordered phase.

\begin{figure}[tb]
\centerline{\includegraphics[width=0.5\textwidth,clip]{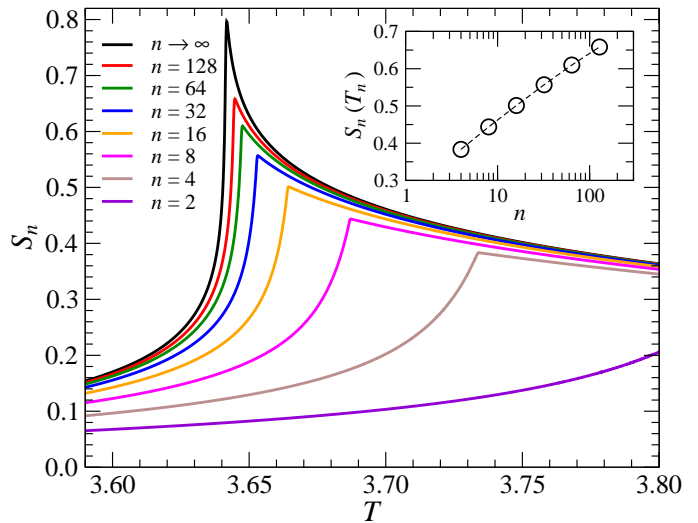}}
\caption{(Color online) Temperature dependence of the entanglement
entropy with respect to $n$.}
\label{fig7}
\end{figure}

As an independent measure of the phase transition, we look at the entanglement entropy 
$S_n^{~}$, which can be directly computed from the reduced density matrix spectrum
\begin{equation}
S_n^{~}( T ) = - 
     {\rm Tr} \left( \rho_{n}^{~} \,  \ln \rho_{n}^{~} \right) \, ,
\label{ent}
\end{equation}
where the reduced density matrices are normalized satisfying the condition
${\rm Tr}\,\rho_n=1$. Figure~\ref{fig7} shows  $S_n^{~}( T )$, where the peak
values, $S_n^{~}( T_n^{~} )$, are shown in the inset. If the curvature radius
$r_n^{~}$ controls the typical length scale at the transition temperature,
it is expected that $S_n^{~}( T_n^{~} )$ behaves as
\begin{equation}
S_n^{~}( T_n^{~} ) \, \sim \, \frac{c}{6} \, \ln \, r_n^{~} \, ,
\end{equation}
where $c$ is the central charge of the system. As shown in Fig.~\ref{fig7},
the increase in $S_n^{~}( T_n^{~} )$ is close to the value $( \ln 2 ) / 12
= 0.05776$ when $n$ doubles, and the fitted value of the slope in the inset
gives $c \sim  0.48$. This value is consistent with $c = 1/2$ in the Ising
universality class. For this reason, our conjecture about the presence of
the typical length scale at $T_n$, which is proportional to $n$
($r_n^{~}\propto 1/n$), is numerically supported.

\section{Conclusions}

We have investigated the thermodynamic property of the Ising model on the
slightly curved $( 3, q_n^{~} )$ lattices, where $q_n^{~}$ represents the
averaged coordination number. We used the CTMRG method to calculate the
thermodynamic functions deep inside the system
around the phase transition temperature. Spontaneous
magnetization suggests a transient behavior from the Ising universality class
to the mean-field one. The specific heat shows a similar transient behavior
from the high temperature side. The entanglement entropy calculated by the
density matrix spectra takes its maximum at the phase transition temperature,
where the peak value is proportional to the logarithm of the curvature radius
of the $( 3, q_n^{~} )$ lattice. These facts support the presence of a finite
thermodynamic length scale at the transition temperature which is proportional
to the curvature radius. Far away from the transition temperature, the
thermal correlation length is much shorter than $n$, therefore, there is
no difference between the flat $( 3, 6 )$ and the $( 3, q_n^{~} )$ lattices 
as long as the thermal property is concerned. As the temperature approaches
the transition temperature, the presence of the length scale prevents the
realization of the criticality without the typical length scale. This could
be the reason of the transient behavior from the Ising universality to the
mean-field one.

\begin{acknowledgments}
A.~G. thanks Frank Verstraete, Sabine Andergassen, and Vladim\'{i}r Bu\v{z}ek
for inspiring discussions and comments. This work was supported by the European
Union projects SIQS, meta-QUTE NFP26240120022, QIMABOS APVV-0808-12, and
VEGA-2/0074/12.  T.~N. and A.~G. acknowledge the support of Grant-in-Aid for
Scientific Research. R.~K. thanks for the support of Japanese Society for the
Promotion of Science P12815.
\end{acknowledgments}

\appendix

\section{Number of sites}

Let us shortly explain another way to count the number of the lattice
sites when the system size $M$ is a multiple of $n$. The extension
process in Eq.~\eqref{tm3qn} describes the whole lattice by the extended
rhombi of the size $n$ by $n$ when the system size $M$ is the multiple
of $n$. In such cases, two lattices with different indices $n$ and $n'$
are mutually similar, where $n$ and $n'$ determine their geometrical
scale. We observe a ``mixed lattice'' under the condition
\begin{equation}
M = n X \, ,
\end{equation}
where $X$ is a positive integer. 

Let us count the number of all lattice points in the corner ${\cal C}_{M}^{~}$
when $M = nX$. We introduce a notation $f^{(X)}_{~}$ for this number. One finds
that there is a recursion relation
\begin{eqnarray}
f^{(X)}_{~} = 
&& 2 \, f^{(X-1)}_{~} - \left[ n( X - 1 ) + 1 \right] \nonumber\\
&& + ( n X + 1 )^2_{~} - \left[ n ( X - 1 ) + 1 \right]^2_{~} \, ,
\end{eqnarray}
where the initial condition is given by
\begin{equation}
f^{(1)}_{~} = ( n + 1 )^2_{~} \, .
\end{equation}
Solving the recursion relation, we have
\begin{equation}
f^{(X)}_{~} = 2^{X-1}_{~} ( 6 n^2_{~} + 2 n ) - ( 2 n^2_{~} - n ) X - 3 n^2_{~} - n + 1 \, .
\end{equation}

In the same manner, we can obtain the number of the special points 
$g^{(X)}_{~}$ inside the corner ${\cal C}_{M = n X}^{~}$, where
the recursion relation in this case can be written as 
\begin{equation}
g^{(X)}_{~} = 2 \, g^{(X-1)}_{~} + 1
\end{equation}
starting with the initial condition $g^{(1)}_{~} = 0$. We get
\begin{equation}
g^{(X)}_{~} = 2^{X-1}_{~} - 1 \, .
\end{equation}

To count the length of the lattice border, we introduce the number of
the border sites $h^{(X)}_{~}$ on the corner $C_{M = nX}^{~}$, where
the border length of the star-shaped region $\left( {\cal C}_{M = nX}^{~}
\right)^6_{~}$ is $6 \, h^{(X)}_{~} - 6$. The recursion relation,
\begin{equation}
h^{(X)}_{~} = 2 \, h^{(X-1)}_{~} - 1 + 2 ( n + 1 ) - 2
\end{equation}
starting from the initial condition
\begin{equation}
h^{(1)}_{~} = 2 n + 1\, ,
\end{equation}
draws the analytic form of the length
\begin{equation}
h^{(X)}_{~} = 2^{X+1}_{~} n - 2 n + 1 \, .
\end{equation}
Expressions for ${\cal N}_n^{~}( M )$, ${\cal S}_n^{~}( M )$,
and ${\cal P}_n^{~}( M )$ for arbitrary $M$ are easily obtained
if one considers the fact that these numbers change polynomially with
respect to $M$ between $M = n X$ and $M = (n+1) X$; the exponential
increase of the lattice sites happens only when the additional corners
are inserted in each $n^{\rm th}$ step.


\begin{thebibliography}{99}

\bibitem{experiment1} H. Yoshikawa, K. Hayashida, Y. Kozuka, A. Horiguchi, and K.
Agawa, Appl. Phys. Lett. {\bf 85}, 5287 (2004).
\bibitem{experiment2} F. Liang, L. Guo, Q.P. Zhong, X.G. Wen, C.P. Chen, N.N.
Zhang, and W.G. Chu, Appl. Phys. Lett. {\bf 89}, 103105 (2006).
\bibitem{experiment3} A. Cabot, A. P. Alivisatos, V. F. Puntes, L. Balcells,
O. Iglesias, and A. Labarta, Phys. Rev. B {\bf 79}, 094419 (2009).
\bibitem{cn} W.A. Moura-Melo, A.R. Pereira, L.A.S. Mol, A.S.T. Pires, Phys.
Lett. A {\bf 360}, 472 (2007).
\bibitem{q-gravity1} V.A. Kazakov, Phys. Lett. A {\bf 119}, 140 (1986).
\bibitem{q-gravity2} C. Holm and W. Janke, Phys. Lett. B {\bf 375}, 69 (1996).
\bibitem{cnet1} D. Krioukov, F. Papadopoulos, A. Vahdat, and M. Bogu\~n\'a, Phys. Rev. E {\bf 80}, 035101 (2009).
\bibitem{cnet2} D. Krioukov, F. Papadopoulos, M. Kitsak, A. Vahdat, and M. Bogu\~n\'a, Phys. Rev. E {\bf 82}, 036106 (2010).
\bibitem{hctmrg-Ising-5-4} K. Ueda, R. Krcmar, A. Gendiar, and T. Nishino, J. Phys.
Soc. Japan {\bf 76}, 084004 (2007).
\bibitem{Shima} H. Shima and Y. Sakaniwa, J. Phys. A {\bf 39}, 4921 (2006).
\bibitem{Sakaniwa} Y. Sakaniwa, H. Shima, Phys. Rev. E {\bf 80}, 021103 (2009)
\bibitem{hctmrg-clock-5-4} A. Gendiar, R. Krcmar, K. Ueda, and T. Nishino, Phys.
Rev. E {\bf 77}, 041123 (2008).
\bibitem{Baek-clock} S.K. Baek, P. Minnhagen, H. Shima, and B.J. Kim, Phys.
Rev. E {\bf 80}, 011133 (2009).
\bibitem{XY-model} S.K. Baek, H. Shima, and B. J. Kim, Phys. Rev. E {\bf 79},
060106(R) (2009).
Academic Press, London, 1982.
\bibitem{hctmrg-J1J2} R. Krcmar, T. Iharagi, A. Gendiar, and T. Nishino, Phys.
Rev. E {\bf 78}, 061119 (2008).
\bibitem{Mosseri} R. Mosseri and J. F. Sadoc, J. Physique Lett. {\bf 43}, 249 (1982).
\bibitem{hctmrg-Ising-3-q} A. Gendiar, R. Krcmar, S. Andergassen, M. Dani\v{s}ka,
and T. Nishino, Phys. Rev. E {\bf 86}, 021105 (2012).
\bibitem{hctmrg-Ising-p-4} R. Krcmar, A. Gendiar, K. Ueda, and T. Nishino, J. Phys.
A {\bf 41}, 125001 (2008).
\bibitem{hctmrg-tr-mat} T. Iharagi, A. Gendiar, H. Ueda, and T. Nishino, J. Phys. Soc. Jpn. {\bf 79}, 104001 (2010).
\bibitem{ctmrg-tn} T. Nishino, J. Phys. Soc. Jpn. {\bf 65}, 891 (1996).
\bibitem{Baxter} R.J.~Baxter, Exactly Solved Models in Statistical Mechanics,
\bibitem{M37} Total number of the vertices in the $( 3, 7)$ lattice geometry, say
$1 + 7 \, \Gamma_M^{~}$, is given by calculating the recurrence relation
$\Gamma_M^{~} = 4 \Gamma_{M-1}^{~} - \Gamma_{M-2}^{~} + 2$ initialized by 
$\Gamma_0^{~} = 0$ and $\Gamma_1^{~} = 2$. It results $\Gamma_M^{~} = 
( \Delta_{+}^{2M+1} - \Delta_{-}^{2M+1} ) / \sqrt{6} - 1$ for which 
$\Delta_\pm^{~} = \sqrt{2 \pm \sqrt{3}}$. Hence, the asymptotic behavior of 
$1 + 7 \, \Gamma_M^{~}$ is proportional to $( 2 + \sqrt{3} )^M_{~}$.
\bibitem{g36} The total number of the sites on the $(3,6)$ lattice is
$\lim_{n\rightarrow\infty}{\cal N}_n(M)=1+12\sum_{j=1}^{M} j \, 2^0_{~} =1+6M(M+1)$.
\end{thebibliography}
\end{document}